\documentclass[12pt]{article}
\usepackage{amsfonts,amssymb,amsmath}
\usepackage{amsthm}
\usepackage{showlabels}
\newcommand\RR{\mathbb R}
\newcommand\CC{\mathbb C}

\newcommand{\sign}{\mathop{\mathrm{sign}}}
\newcommand\beq{\begin{equation}}
\newcommand\eeq{\end{equation}}
\newtheorem{theorem}{Theorem}

\begin{document}
\title{Creation and annihilation of point-potentials using Moutard-type transform in spectral variable\thanks{The main part of the work was fulfilled during the visit of the 
first author to the IHES, France in November 2019. The work was partially supported by a joint grant of the Russian Foundation for Basic Research and CNRS (project no. RFBR 17-51-150001 NCNI\_a/PRC 1545 CNRS/RFBR).}}

\author{P.G. Grinevich
\thanks{Steklov Mathematical Institute of Russian Academy of Sciences, Moscow, 199911, Russia;
L.D. Landau Institute for Theoretical Physics, Chernogolovka, 142432,
Russia; Lomonosov Moscow State University, Moscow, 119991, Russia; e-mail: pgg@landau.ac.ru} \and R.G. Novikov\thanks
{CNRS (UMR 7641), Centre de Math\'ematiques Appliqu\'ees, 
\'Ecole Polytechnique, 91128, Palaiseau, France;
IEPT RAS, 117997, Moscow, Russia;
e-mail: novikov@cmap.polytechnique.fr}}
\date{}
\maketitle

\begin{abstract}
We continue to develop the method for creation and annihilation of contour singularities in the $\bar\partial$--spectral data for the two-dimensional Schr\"odinger equation at fixed energy. Our method is based on the Moutard-type transforms for generalized analytic functions. In this note we show that this approach successfully works for point potentials.
\end{abstract}

Key words: two-dimensional Schr\"odinger equation, Faddeev eigenfunctions, fixed-energy spectral problem, spectral singularities, generalized analytic functions, Moutard-type transforms, point potentials.

\medskip

MSC: 30G20, 37K35, 35J10, 47A40, 81Q05, 81U15. 

\medskip

\section{Introduction} 

We continue studies of the two dimensional Schr\"odinger equation at fixed energy $E$
\beq
\label{eq:0.1}
-\Delta\psi + v(x)\psi = E\psi,  \ \ x\in\RR^2,
\eeq
where $v(x)$ is a real-valued function on $\RR^2$ with sufficient decay at infinity. 

More precisely, for equation (\ref{eq:0.1}) we continue studies of the direct and inverse scattering transform. 
We are focused on the simplest case of real negative energy $E$.

In order to define generalized scattering data $B=B_E$ for equation (\ref{eq:0.1}) at a negative energy $E$, one uses the Faddeev's eigenfunctions 
$\psi=\psi(x,k)$ (see \cite{F1,F2,N,G,NKh}), specified by the asymptotic condition:
\beq
\label{eq:0.2}
\psi= e^{ikx}\left(1+o(1) \right) \ \
\mbox{as} \ \ |x|\rightarrow\infty, \ \ k\in\Sigma_E,
\eeq
where
\beq
\label{eq:0.3}
\Sigma_E=\{k\in\CC^2:k^2=E \}, \ \ E<0, \ \  E\in\RR.
\eeq
We say that $\Sigma_E$ is the spectral variety for fixed $E$. We recall that $\Sigma_E\approx\CC\backslash0$ and has the following parametrization: 
\beq
\label{eq:0.4}
\Sigma_E= \left\{k_E(\lambda)=\left(\left(\lambda +\frac{1}{\lambda} \right)\frac{i\sqrt{|E|}}{2}, 
\left(\lambda -\frac{1}{\lambda} \right)\frac{\sqrt{|E|}}{2}    \right)
:\lambda\in\CC\backslash0 \right\}.
\eeq
We recall that the restriction of the Faddeev eigenfunction $\psi(x,k_E(\lambda))$ to $\Sigma_E$ for each fixed $x$ is a generalized analytic function with respect to the spectral variable $\lambda$; see \cite{GN,G}:
\beq
\frac{\partial}{\partial\bar\lambda}\psi(x,k_E(\lambda)) = B_E(\lambda)\overline{\psi(x,k_E(\lambda))}, \ \ E<0, \ \ \lambda\in\CC\backslash0,
\label{eq:0.5} 
\eeq
and
\beq
\label{eq:0.6}
\psi= e^{ik_E(\lambda)x}\left(1+o(1) \right)\ \ 
\mbox{as} \ \ |\lambda| \rightarrow\infty \ \ \mbox{and as} \ \ |\lambda| \rightarrow0.
\eeq

The coefficient $B_E(\lambda)$ in (\ref{eq:0.5}) does not depend on $x$, and plays the role of generalized scattering data ($\bar\partial$--spectral data) for equation (\ref{eq:0.1}).  Here and below, the notation $f=f(\lambda)$ does not mean that $f$ is holomorphic in $\lambda$. 

For an introduction to the classical theory of generalized analytic functions (pseudoanalytic functions), see \cite{V,Bers}.

Note that for real $v(x)$ the data $B_E$ has the following symmetries (see \cite{GN}):
\beq
\label{eq:0.7}
B\left(\frac{1}{\bar \lambda} \right)= -\lambda^2\, \overline{B(\lambda)}, \ \ B\left(-\frac{1}{\bar \lambda}\right) = \lambda\bar\lambda\ B(\lambda), \ \
\lambda\in\CC\backslash0.
\eeq

We consider the following inverse scattering problem.

\textit{Problem 1.} Given $B_E(\lambda)$ at fixed $E<0$, find $v(x)$. 

Note that Problem~1 has direct applications to integrating the Novikov-Veselov hierarchy; see, for example, \cite{VN,G,GN}. Note also that Problem~1 admits direct applications to reconstructing potential $v$, supported in a bounded domain, from the Dirichlet--to--Newman boundary map; see, for example, \cite{N2,N3}.

The standard approach to Problem~1 uses the following scheme:
\beq
\label{eq:1.7.2}
B_E\xrightarrow{(\ref{eq:0.5}),(\ref{eq:0.6})} \psi \xrightarrow{(\ref{eq:0.1})} v.
\eeq

It is well-known that Problem~1 is uniquely solvable and scheme (\ref{eq:1.7.2}) works perfectly if $B$ satisfies (\ref{eq:0.7}), and $B\in L_p$, $p>2$, at the unit disc $|\lambda|\le1$ (see \cite{GN,N}). These conditions are fulfilled if $v(x)$ is sufficiently regular and $|E|$ is sufficiently large. However, if $E_0<E<0$, where $E_0$ denotes the smallest discrete eigenvalue (ground state energy), then $B$ and $\psi$ have strong singularities, which are typically contour poles, and the classical methods of the generalized analytic functions theory do not work. In \cite{GN} hypothetical solubility conditions on the singularities at the contours were formulated. In \cite{GRN2,GRN3,GRN4} we have shown that singularities with these constraints can be locally removed by applying the Moutard-type transformations for generalized analytic functions. In turn, the construction of  \cite{GRN2,GRN3,GRN4} was stimulated by articles \cite{T1,T2} by I.A. Taimanov dedicated to the Moutard-type transforms for the Dirac operators. For extra information about Motard-type transforms, see also \cite{Mout,TT,MatvSal,YLW,NS,DGNS}.

In this note for the first time we show that, at least in some interesting cases, our approach to creation and annihilation of spectral contour singularities works globally. 
In particular, we demonstrate global creation and annihilation of spectral contour singularities for the case of two-dimensional analogs of the Bethe-Peierls-Fermi-Zel'dovich-Berezin-Faddeev point potentials.

As about other approaches to fixed-energy inverse problems for the Schr\"odinger equation (\ref{eq:0.1}) in the case when Faddeev's spectral singularities may be present, see, for example, \cite{N,Buck,LNV}.

\section{``Scattering'' functions} 

We recall that for sufficiently regular $v(x)$ the Faddeev eigenfunctions $\psi(x,k)$ satisfy the following generalized Lippmann-Schwinger integral equation:
\beq
\label{eq:1.5.1}
\psi(x,k) =  e^{ikx} + \iint_{\RR^2} G(x-y,k) v(y) \psi(y,k) dy,
\eeq
where
\beq
\label{eq:1.5.2}
G(x,k) =  e^{ikx} g(x,k),
\eeq
\beq
\label{eq:1.5.3}
 g(x,k)=-\left(\frac{1}{2\pi}\right)^2\iint_{\RR^2} 
\frac{e^{i\xi x}}{\xi^2+2k\xi}  d\xi, \ \ k\in\Sigma_{E}, \ \ x\in\RR^2, \ \ E<0.
\eeq
Here $\Sigma_{E}$ is defined by (\ref{eq:0.4}), $G(x,k)$ is  the Faddeev's Green function for $\Delta + k^2$. Note also that 
\beq
\label{eq:1.5}
|\Re k_E(\lambda)|+ |\Im k_E(\lambda)| = \left\{\begin{aligned} 
& \sqrt{|E|} |\lambda|, &  |\lambda|\ge 1,\\
& \sqrt{|E|} |\lambda|^{-1}, &  |\lambda|< 1,
 \end{aligned} \right. \ \ E<0,
\eeq
where $k_{E}$ is defined in (\ref{eq:0.4}).

In addition, the following formulas hold:

\beq
\label{eq:1.7.1} 
B_E(\lambda)  =
\frac{\pi\sign(\lambda\bar\lambda-1)}{\bar\lambda} b(k_E(\lambda)),
\eeq
where 
\beq
\label{eq:1.8.1}
b(k) =\frac{1}{(2\pi)^2}\iint_{\RR^2} e^{i\bar k x} v(x) \psi(x,k)dx.
\eeq
Here $B_E$ is the $\bar\partial$--spectral data arising in (\ref{eq:0.5}),  $b(\lambda)= b_E(\lambda) = b(k_E(\lambda))$ is a part of the Faddeev generalized scattering data. Actually, $b(\lambda)$ can be treated as a non-linear analog of the Fourier transform of the potential $v(x)$. For real $v(x)$ the function $b(\lambda)$ has the following symmetries \cite{GN}:
\beq
\label{eq:1.8}
b\left(\frac{1}{\bar \lambda} \right)= \overline{b(\lambda)}, \ \ b\left(-\frac{1}{\bar \lambda}\right) = b(\lambda), \ \
\lambda\in\CC\backslash0.
\eeq
Symmetries (\ref{eq:1.8}) are equivalent to symmetries (\ref{eq:0.7}).

If the potential $v(x)$ depends only of the distance from the origin: $v(x)=v(|x|)$, then 
\beq
\label{eq:1.9}
b(\lambda) = b(|\lambda|), \ \ b(\lambda)=\overline{b(\lambda)}, \ \ \lambda\in\CC\backslash0.
\eeq

\section{Point potentials} 
\label{sec:ptp}

For the two-dimensional analogs of the Bethe-Peierls-Fermi-Zel'dovich-Berezin-Faddeev point potentials $v_{0,\alpha}(x)$ with support at the 
point $x=\{0\}$, the following formulas hold (see \cite{GRN4}): 
\beq
\label{eq:2.1}
\psi = \psi_{0,\alpha}(x,k)=e^{ikx}\left[
1 + \frac{\alpha}{1-\frac{\alpha} {2\pi}\ln(|\Re k| + |\Im k|) }\cdot g(x,k)\right],
\eeq
\beq
\label{eq:2.2}
b_E = b_{0,\alpha}(k) = \left(\frac{1}{2\pi}\right)^2  \frac{\alpha}{1-\frac{\alpha}
{2\pi}\ln(|\Re k| + |\Im k|)},
\eeq
where $g(x,k)$ is defined in (\ref{eq:1.5.3}), $\alpha\in\RR$, $k\in\Sigma_{E}$, $x\in\RR^2$, $E<0$. 

We recall that the function $\psi_{0,\alpha}(x,k)$ satisfies the Schr\"odinger equation (\ref{eq:0.1}) with the point-type potential $v(x)=v_{0,\alpha}(x)$  and condition (\ref{eq:0.2}) at infinity (the exact meaning of the Schr\"odinger equation with the point-type potential is discussed, in particular, in \cite{AGHH}, \cite{GRN}).

In addition, the $\bar\partial$-equation (\ref{eq:0.5}) holds, where $\psi$, $b$ are given by (\ref{eq:2.1}), (\ref{eq:2.2}) (see \cite{GRN}). Note also that 
\beq
\label{eq:2.3}
\frac{\partial}{\partial\bar\lambda} a_{0,\alpha}(k_E(\lambda))  = B(\lambda) \overline{a_{0,\alpha}(k_E(\lambda))},
\ \ E<0, \ \ \lambda\in\CC\backslash 0,
\eeq
where  $B$ is given by (\ref{eq:1.7.1}), (\ref{eq:2.2}) and
\beq
\label{eq:2.4}
a_{0,\alpha}(k)  = \lim\limits_{|x|\rightarrow 0}\frac{1}{2\pi}\big(\ln(|x|)\big)^{-1} \psi_{0,\alpha}(x,k),
\eeq
\beq
\label{eq:2.5}
a_{0,\alpha}(k) = \overline{a_{0,\alpha}(k)} = b_{0,\alpha}(k), \ \ k\in\Sigma_E, \ \ E<0.
\eeq

\section{Moutart-type transforms for generalized analytic functions}

Consider the basic pair of conjugate equations of the generalized analytic function theory:  
\begin{align}
\label{eq:3.1}
&\partial_{\bar\lambda} \psi = B \bar \psi \ \ \mbox{in} \ \ D,\\
\label{eq:3.2}
&\partial_{\bar\lambda} \psi^* = -\bar B \bar \psi^* \ \ \mbox{in} \ \ D,
\end{align}
where $D$ is an open domain in $\CC$,  $B=B(\lambda)$ is a given function in $D$, 
$\partial_{\bar \lambda}=\partial/\partial\bar \lambda$; see \cite{V}.

Let $f(\lambda)$, $f^*(\lambda)$ denote a pair of fixed solutions of equations (\ref{eq:3.1}) and (\ref{eq:3.2}), respectively.
A simple Moutard-type transform $\mathcal{M}=\mathcal{M}_{B,f,f^*}$ for the pair of conjugate equations (\ref{eq:3.1}), (\ref{eq:3.2}) is given by 
the formulas (see \cite{GRN}):
\beq
\label{eq:3.3}
\tilde B = \mathcal{M} B= B + \frac{f\overline{f^*}}{\omega_{f,f^*}},
\eeq
\beq
\label{eq:3.4}
\tilde\psi=\mathcal{M} \psi=
\psi-\frac{\omega_{_{\psi,f^*}}}{\omega_{f,f^*}}\,f , \ \ 
\tilde{\psi}^*=  \mathcal{M} \psi^*=  \psi^* - \frac{\omega_{f,\psi^*}}{\omega_{f,f^*}}\, f^*,
\eeq
where $\psi$ and  $\psi^*$ are arbitrary solutions of (\ref{eq:3.1}) and (\ref{eq:3.2}), respectively, and  $\omega_{\psi,\psi^*}=\omega_{\psi,\psi^*}(\lambda)$ denotes 
imaginary-valued function defined by:
\beq
\label{eq:3.5}
\partial_{\lambda} \omega_{\psi,\psi^*} =\psi\psi^*, \ \ 
\partial_{\bar\lambda} \omega_{\psi,\psi^*} =-\overline{\psi\psi^*} \ \ \mbox{in} \ \ D,
\eeq
where this definition is self-consistent, at least, for simply connected $D$, 
whereas a pure imaginary integration constant may depend on concrete 
situation. The point is that the functions $\tilde\psi$, $\tilde\psi^*$ defined in (\ref{eq:3.4}) 
satisfy the conjugate pair of Moutard-transformed equations:
\begin{align}
\label{eq:3.6}
&\partial_{\bar\lambda} \tilde\psi= \tilde B\, \overline{\tilde\psi}& \ \ &\mbox{in} \ \ D, \\ 
\label{eq:3.7}
&\partial_{\bar\lambda} \tilde\psi^*= -\overline{\tilde B}\, \overline{\tilde\psi^*}& \ \ 
&\mbox{in} \ \ D.
\end{align}

\section{Creation of point potentials by Moutard-type transforms}

Let us start with the Schr\"odinger equation (\ref{eq:0.1}) with the zero potential $v(x)\equiv0$. In this case for $E<0$ one has:
\beq
\label{eq:4.1}
\psi(x,k) = e^{ikx} = e^{-\frac{\sqrt{|E|}}{2}\left(\lambda\bar z + \frac{z}{\lambda} \right)}, \ \ \lambda\in\CC\backslash0, 
\eeq
\beq
\label{eq:4.2}
B(\lambda)=B_E(\lambda) \equiv0,
\eeq
where $B(\lambda)$ is defined in (\ref{eq:1.7.1}),
\beq
\label{eq:4.3}
z = x_1 + i x_2, \ \ \bar z = x_1 - i x_2, \ \ \partial_z = \frac{1}{2}(\partial_{x_1}-i\partial_{x_2}), \ \ 
\partial_{\bar z} = \frac{1}{2}(\partial_{x_1} + i\partial_{x_2}), 
\eeq
and $\lambda$ is the same as in (\ref{eq:0.4}). 

Let  $D_+$ and  $D_-$ denote the set of points inside and outside the unit circle respectively:
\beq
\label{eq:4.4}
\begin{split}
D_+ = \{\lambda \in\CC:  0<|\lambda| \le 1 \}\\
D_- = \{\lambda \in\CC: |\lambda| \ge 1 \}
\end{split}
\eeq

Equations (\ref{eq:3.1}), (\ref{eq:3.2}) with $B\equiv0$ have the following pair of solutions
\beq
\label{eq:4.5}
f(\lambda) = 1, \ \ f^*(\lambda) = \frac{i \sign(\lambda\bar\lambda-1)}{\lambda}, \ \ \mbox{for} \ \ \lambda\in D_+ \ \ \mbox{or} \ \ \lambda\in D_-.
\eeq

For the conjugate equation (\ref{eq:3.2}) with $B\equiv0$ we also consider the solution
\beq
\label{eq:4.5.1}
\psi^*(z,\lambda,E) = \frac{i}{\lambda}\exp{\left[\frac{\sqrt{|E|}}{2}\left(\lambda\bar z + \frac{z}{\lambda}  \right) \right]}, \ \ E<0, \ \ \lambda\in\CC\backslash0.
\eeq

Consider Moutard-type transforms $\mathcal{M}^{\pm}=\mathcal{M}^{\pm}_{B,f,f^*}$ for equations (\ref{eq:3.1}), (\ref{eq:3.2}) in the domains $D_{\pm}$, respectively, constructed according to formulas (\ref{eq:3.3})-(\ref{eq:3.5}). 

\begin{theorem} \label{thm:1}
Let $\tilde B_{\pm} = \mathcal{M}^{\pm}_{B,f,f^*} B$, $\tilde \psi_{\pm} = \mathcal{M}^{\pm}_{B,f,f^*} \psi$, $\tilde\psi^{*}_{\pm} = \mathcal{M}^{\pm}_{B,f,f^*} \psi^{*}$ in $D_{\pm}$, where $B$, $\psi$, $\psi^*$ are given by (\ref{eq:4.2}),  (\ref{eq:4.1}), (\ref{eq:4.5.1}) (and correspond to the Schr\"odinger equation (\ref{eq:0.1}), with $v(x)\equiv0$, $E<0$), $f(\lambda)$, $f^*(\lambda)$ are given by (\ref{eq:4.5}).

Then for a proper choice of integration constants in formulas (\ref{eq:3.5}) we obtain:
\beq
\label{eq:4.6}
\tilde B_{\pm}(\lambda)= \tilde B(\lambda) = \frac{\pi\sign(\lambda\bar\lambda-1)}{\bar\lambda}b_{0,\alpha}(k_{E}(\lambda)),
\eeq
\beq
\label{eq:4.7}
\tilde\psi_{\pm}(z,\lambda)=\tilde \psi(z,\lambda) = \psi_{0,\alpha}(x,k_{E}(\lambda)),
\eeq
\beq
\label{eq:4.7.1}
\tilde\psi^*_{\pm}(z,\lambda) =\tilde \psi^*(z,\lambda)  = \frac{i}{\lambda} \psi_{0,\alpha}(x,k_{E}(-\lambda)),
\eeq
where $\lambda\in\CC\backslash0$, $b_{0,\alpha}$, $\psi_{0,\alpha}$ are given by (\ref{eq:2.2}), (\ref{eq:2.1}) (and correspond to the point potential with the support at the point $\{0\}$).
\end{theorem}
\textit{Proof of Theorem~\ref{thm:1}.} Under our assumptions, for potentials $\omega_{f,f^*}(\lambda)$, $\omega_{\psi,f^*}(\lambda)$ in
(\ref{eq:3.3})-(\ref{eq:3.5}) the following formulas are valid:
\beq
\label{eq:4.8}
\omega_{f,f^*}(\lambda) = i [|\ln(\lambda\bar\lambda)|+\ln(|E|)]  - i\left\{\begin{array}{ll} 
c^+_{f,f^*}, & \lambda \in D_+\\
c^-_{f,f^*},  & \lambda \in D_-,
\end{array} \right.
\eeq
\beq
\label{eq:4.9}
\omega_{\psi,f^*}(\lambda) = 4 \pi i G(z,\lambda,E) +  i \left\{\begin{array}{ll} 
c^{+}_{\psi,f^*}(z),   & \lambda \in D_+\\
c^{-}_{\psi,f^*}(z),   & \lambda \in D_-,
\end{array} \right.
\eeq
\beq
\label{eq:4.9.1}
\omega_{f,\psi^*} = 4\pi i \sign(\lambda\bar\lambda-1)G(z,-\lambda,E) + i
\left\{\begin{array}{ll} 
c^{+}_{f,\psi^*}(z),   & \lambda \in D_+\\
c^{-}_{f,\psi^*}(z),   & \lambda \in D_-,
\end{array} \right. 
\eeq
where $c^{+}_{f,f^*}$, $c^{-}_{f,f^*}$, $c^{+}_{\psi,f^*}(z)$, $c^{-}_{\psi,f^*}(z)$, $c^{+}_{f,\psi^*}(z)$, $c^{-}_{f,\psi^*}(z)$ are some real constants with respect to  $\lambda$, $G(z,\lambda,E)=G(x,k_{E}(\lambda))$ is defined by (\ref{eq:1.5.2}), (\ref{eq:1.5.3}).
Formula (\ref{eq:4.8}) can be checked directly. Formulas  (\ref{eq:4.9}), (\ref{eq:4.9.1}) can be also checked directly using the relations
\beq
\label{eq:4.10}
\frac{\partial}{\partial\bar\lambda} G(z,\lambda,E) =\frac{1}{4\pi}\frac{\sign(\lambda\bar\lambda-1)}{\bar\lambda}
\exp{\left[-\frac{\sqrt{|E|}}{2}\left(\bar\lambda z + \frac{\bar z}{\bar\lambda}  \right) \right]},
\eeq
\beq
\label{eq:4.11}
\overline{G(z,\lambda,E)} = G(z,\lambda,E), \ \ \mbox{where} \ \ \lambda\in\CC\backslash0,
\eeq
(see \cite{N}, page 423).

Using formulas (\ref{eq:3.3}), (\ref{eq:4.2}), (\ref{eq:4.5}) and (\ref{eq:4.8}) with $c^{+}_{f,f^*} = c^{-}_{f,f^*}= \frac{4\pi}{\alpha}$, we obtain formula (\ref{eq:4.6}). Using formulas (\ref{eq:3.4}), (\ref{eq:4.1}), (\ref{eq:4.5}), (\ref{eq:4.5.1}) and (\ref{eq:4.9}), (\ref{eq:4.9.1}) with $c^{+}_{\psi,f^*}(z)=c^{-}_{\psi,f^*}(z)\equiv0$, and $c^{+}_{f,\psi^*}(z)=c^{-}_{f,\psi^*}(z)  \equiv0$ we obtain the formulas:
 \beq
\label{eq:4.12}
\begin{split}
\tilde\psi_{\pm}(z,\lambda) =\tilde \psi(z,\lambda)  =  \exp{\left[-\frac{\sqrt{|E|}}{2}\left(\lambda\bar z + \frac{z}{\lambda} \right)\right]} -\frac{4\pi\, G(z,\lambda,E)}{|\ln(\lambda\bar\lambda)|+\ln(|E|)-4\pi/\alpha},
\end{split}
\eeq
\beq
\label{eq:4.12.1}
\tilde\psi^*_{\pm}(z,\lambda) =\tilde \psi^*(z,\lambda)  =  \frac{i}{\lambda}\left\{\exp{\left[\frac{\sqrt{|E|}}{2}\left(\lambda\bar z + \frac{z}{\lambda}  \right)\right]} -\frac{4\pi\, G(z,-\lambda,E)}{|\ln(\lambda\bar\lambda)|+\ln(|E|)-4\pi/\alpha}    \right\},
\eeq
where $\lambda\in\CC\backslash0$. Formulas (\ref{eq:4.12}), (\ref{eq:4.12.1}) can be rewritten as (\ref{eq:4.7}), (\ref{eq:4.7.1}).
\qed

\section{Annihilation of point potentials by Moutard-type transforms}

Now we start with the Schr\"odinger equation (\ref{eq:0.1}) with the point-type potential $v_{0,\alpha}(x)$ mentioned in the Section~\ref{sec:ptp} . In this case for $E<0$ one has:
\beq
\label{eq:5.1}
B(\lambda) = B_E(\lambda) = -\frac{\sign(\lambda\bar\lambda-1)}{\bar\lambda}\,\frac{1}{|\ln(\lambda\bar\lambda)|+\ln(|E|)-4\pi/\alpha},
\eeq
\beq
\label{eq:5.1.1}
\begin{split}
\psi(z,\lambda)  = \psi(z,\lambda,E) = \exp{\left[-\frac{\sqrt{|E|}}{2}\left(\lambda\bar z + \frac{z}{\lambda} \right)\right]} -\frac{4\pi\, G(z,\lambda,E)}{|\ln(\lambda\bar\lambda)|+\ln(|E|)-4\pi/\alpha},
\end{split}
\eeq
\beq
\label{eq:5.1.2}
\psi^*(z,\lambda) = \psi^*(z,\lambda,E) =  \frac{i}{\lambda}\left\{\exp{\left[\frac{\sqrt{|E|}}{2}\left(\lambda\bar z + \frac{z}{\lambda}  \right)\right]} -\frac{4\pi\, G(z,-\lambda,E)}{|\ln(\lambda\bar\lambda)|+\ln(|E|)-4\pi/\alpha}    \right\},
\eeq
and $\psi$,  $\psi^*$ satisfy the Schr\"odinger equation (\ref{eq:0.1}) with $v(x)=v_{0,\alpha}(x)$ and the $\bar\partial$-equations (\ref{eq:3.1}),  (\ref{eq:3.2}),  where $B$ is given by (\ref{eq:5.1}), $D=D_{\pm}$. 

In addition, equations (\ref{eq:3.1}),  (\ref{eq:3.2}),  where $B$ is given by (\ref{eq:5.1}), $D=D_{\pm}$,  have the following particular pair of conjugate solutions:
\beq
\label{eq:5.2}
f = a(\lambda) = -\frac{1}{\pi}\,\frac{1}{|\ln(\lambda\bar\lambda)|+\ln(|E|)-4\pi/\alpha},
\eeq
\beq
\label{eq:5.3}
f^* = \frac{i\sign(\lambda\bar\lambda-1)}{\lambda}\,a(\lambda) = -\frac{i\sign(\lambda\bar\lambda-1)}{\pi\lambda}\,\frac{1}{|\ln(\lambda\bar\lambda)|+\ln(|E|)-4\pi/\alpha}.
\eeq
Here $a(\lambda) = a_E(\lambda) =a_{0,\alpha}(\lambda)$ is given by (\ref{eq:2.4}), (\ref{eq:2.5}).

\begin{theorem}\label{thm:2}
Let $\tilde B_{\pm} = \mathcal{M}^{\pm}_{B,f,f^*} B$, $\tilde \psi_{\pm} = \mathcal{M}^{\pm}_{B,f,f^*} \psi$, $\tilde\psi^{*}_{\pm} = \mathcal{M}^{\pm}_{B,f,f^*} \psi^{*}$ in $D_{\pm}$, where $B$, $\psi$, $\psi^*$ are given by (\ref{eq:5.1}),  (\ref{eq:5.1.1}), (\ref{eq:5.1.2}) (and correspond to the Schr\"odinger equation (\ref{eq:0.1}), with $v(x)=v_{0,\alpha}(x)$, $E<0$), $f(\lambda)$, $f^*(\lambda)$ are given by (\ref{eq:5.2}), (\ref{eq:5.3}).

Then for a proper choice of integration constants in formulas (\ref{eq:3.5}) we obtain:
\beq
\label{eq:5.4}
\tilde B(\lambda)\equiv 0,
\eeq
\beq
\label{eq:5.5}
\tilde\psi_{\pm}(z,\lambda,E) =\tilde\psi(z,\lambda,E)=\exp{\left[-\frac{\sqrt{|E|}}{2}\left(\lambda\bar z + \frac{z}{\lambda} \right)\right]},
\eeq
\beq
\label{eq:5.6}
\tilde\psi^*_{\pm}(z,\lambda,E) =\tilde\psi^*(z,\lambda,E)=\frac{i}{\lambda}\exp{\left[\frac{\sqrt{|E|}}{2}\left(\lambda\bar z + \frac{z}{\lambda} \right)\right]},
\eeq
where $\lambda\in\CC\backslash0$. 
\end{theorem}

The point is that the Moutard-transformed data $\tilde B$  and eigenfucntions $\tilde\psi$, $\tilde\psi^*$ correspond to the Schr\"odinger equation (\ref{eq:0.1}) with zero potential $v(x)\equiv0$ at fixed energy $E<0$.

\textit{Proof of Theorem~\ref{thm:2}.}  Under our assumptions, for potentials $\omega_{f,f^*}(\lambda)$, $\omega_{\psi,f^*}(\lambda)$ in (\ref{eq:3.3})-(\ref{eq:3.5}) the following formulas are valid:
\beq
\label{eq:5.7}
\omega_{f,f^*}(\lambda) =\frac{i}{\pi}a(\lambda) + \left\{\begin{array}{ll} 
c^{+}_{f,f^*}, & \lambda \in D_+\\
c^{-}_{f,f^*}, & \lambda \in D_-,
\end{array}\right.
\eeq
\beq
\label{eq:5.8}
\omega_{\psi,f^*}=-\frac{4i\, G(z,\lambda,E)}{|\ln(\lambda\bar\lambda)|+\ln(|E|)-4\pi/\alpha} + i 
\left\{\begin{array}{ll}
c^{+}_{\psi,f^*}(z),  & \lambda \in D_+\\
c^{-}_{\psi,f^*}(z),  & \lambda \in D_-,
\end{array}\right.
\eeq
\beq
\label{eq:5.9}
\omega_{f,\psi^*}=-\frac{4i\, \sign(\lambda\bar\lambda-1)\,G(z,-\lambda,E)}{|\ln(\lambda\bar\lambda)|+\ln(|E|)-4\pi/\alpha}  + i 
\left\{\begin{array}{ll}
c^{+}_{f,\psi^*}(z),  & \lambda \in D_+\\
c^{-}_{f,\psi^*}(z),  & \lambda \in D_-,
\end{array}\right.
\eeq
where $c^{+}_{f,f^*}$, $c^{-}_{f,f^*}$, $c^{+}_{\psi,f^*}(z)$, $c^{-}_{\psi,f^*}(z)$, $c^{+}_{f,\psi^*}(z)$, $c^{-}_{f,\psi^*}(z)$ are some real constants with respect to  $\lambda$, $G(z,\lambda,E)=G(x,k_{E}(\lambda))$ is defined by (\ref{eq:1.5.2}), (\ref{eq:1.5.3}).

Formula (\ref{eq:5.7}) follows from the formulas
\beq
\label{eq:5.10}
\frac{\partial}{\partial\bar\lambda} a(\lambda) = \frac{\pi\sign(\lambda\bar\lambda-1)}{\bar\lambda}\big(a(\lambda)\big)^2,
\eeq
and
\beq
\label{eq:5.11}
\frac{\sign(\lambda\bar\lambda-1)}{\bar\lambda}\big(a(\lambda)\big)^2= i \overline{f(\lambda) f^*(\lambda)}.
\eeq
Formula (\ref{eq:5.10}) follows from the formulas (\ref{eq:2.3}), (\ref{eq:2.5}). Finally, formula (\ref{eq:5.11}) follows from (\ref{eq:5.2}),  (\ref{eq:5.3}).

To prove (\ref{eq:5.8}), we use the formulas:
\beq
\label{eq:5.12}
\begin{split}
\psi(z,\lambda) - \exp{\left[-\frac{\sqrt{|E|}}{2}\left(\lambda\bar z + \frac{z}{\lambda} \right)\right]} =-\frac{4\pi\, G(z,\lambda,E)}{|\ln(\lambda\bar\lambda)|+\ln(|E|)-4\pi/\alpha},
\end{split}
\eeq
\beq
\label{eq:5.13}
\begin{split}
\frac{\partial}{\partial\bar\lambda} \psi(z,\lambda) = B(\lambda)\overline{\psi(z,\lambda)}
= i\pi \overline{\psi(z,\lambda)\,f^*(\lambda)},
\end{split}
\eeq
and formula (\ref{eq:4.11}).

To prove (\ref{eq:5.9}), we use the formulas :
\beq
\label{eq:5.14}
\psi^*(z,\lambda) - \frac{i}{\lambda}\exp{\left[\frac{\sqrt{|E|}}{2}\left(\lambda\bar z + \frac{z}{\lambda}  \right)\right]} =-\frac{i}{\lambda}\frac{4\pi\, G(z,-\lambda,E)}{|\ln(\lambda\bar\lambda)|+\ln(|E|)-4\pi/\alpha}  ,
\eeq
\beq
\label{eq:5.15}
\begin{split}
\frac{\partial}{\partial\bar\lambda} \frac{\lambda}{\pi}\psi^*(z,\lambda) = \frac{\lambda}{\pi}\,
\overline{B(\lambda)}\overline{\psi^*(z,\lambda)}=\sign(\lambda\bar\lambda-1) a(\lambda) \overline{\psi^*(z,\lambda)}=\\
=\sign(\lambda\bar\lambda-1) \overline{f(\lambda)} \overline{\psi^*(z,\lambda)},
\end{split}
\eeq
and formula (\ref{eq:4.11}).

Next, we choose 
\beq
\label{eq:5.16}
c^{+}_{f,f^*} = c^{-}_{f,f^*}=0,\ \  c^{+}_{\psi,f^*}(z)=c^{-}_{\psi,f^*}(z) = c^{+}_{f,\psi^*}(z)=c^{-}_{f,\psi^*}(z)  \equiv0.
\eeq

Formula (\ref{eq:5.4}) follows from the following computation:
\begin{align}
\tilde B(\lambda) &=\frac{\pi\sign(\lambda\bar\lambda-1)}{\bar\lambda} a(\lambda)+ \frac{f\overline{f^*}}{\omega_{f,f^*}}=
\nonumber\\
&=\frac{\pi\sign(\lambda\bar\lambda-1)}{\bar\lambda} a(\lambda)+ \frac{a(\lambda)\frac{-i\sign(\lambda\bar\lambda-1)}{\bar\lambda}\,a(\lambda) }{\frac{i}{\pi}a(\lambda)  }= \label{eq:5.20} \\
&=\frac{\pi\sign(\lambda\bar\lambda-1)}{\bar\lambda} a(\lambda)+ a(\lambda)\frac{-\pi\sign(\lambda\bar\lambda-1)}{\bar\lambda}=0.\nonumber
\end{align}
Formula (\ref{eq:5.5}) follows from the following computation:
\begin{align}
\tilde \psi(z,\lambda) &=\psi(z,\lambda) - \frac{\omega_{\psi,f^*}}{\omega_{f,f^*}}f = \nonumber  \\
&=  \exp{\left[-\frac{\sqrt{|E|}}{2}\left(\lambda\bar z + \frac{z}{\lambda} \right)\right]} -\frac{4\pi\, G(z,\lambda,E)}{|\ln(\lambda\bar\lambda)|+\ln(|E|)-4\pi/\alpha}+\nonumber\\
& +\frac{4i\, G(z,\lambda,E)}{|\ln(\lambda\bar\lambda)|+\ln(|E|)-4\pi/\alpha}\frac{a(\lambda)}{\frac{i}{\pi}a(\lambda) }=
\label{eq:5.21}\\
&=\exp{\left[-\frac{\sqrt{|E|}}{2}\left(\lambda\bar z + \frac{z}{\lambda} \right)\right]} -\frac{4\pi\, G(z,\lambda,E)}{|\ln(\lambda\bar\lambda)|+\ln(|E|)-4\pi/\alpha}+ \nonumber\\
& +\frac{4\pi\, G(z,\lambda,E)}{|\ln(\lambda\bar\lambda)|+\ln(|E|)-4\pi/\alpha}\frac{a(\lambda)}{a(\lambda) }= \nonumber\\
&=\exp{\left[-\frac{\sqrt{|E|}}{2}\left(\lambda\bar z + \frac{z}{\lambda} \right)\right]}. \nonumber
\end{align}
Formula (\ref{eq:5.6}) follows from the following computation:
\begin{align}
\tilde \psi^*(z,\lambda) &=\psi^*(z,\lambda) - \frac{\omega_{f,\psi^*}}{\omega_{f,f^*}}f^* = \nonumber \\
&=\frac{i}{\lambda}\left\{\exp{\left[\frac{\sqrt{|E|}}{2}\left(\lambda\bar z + \frac{z}{\lambda}  \right)\right]} -\frac{4\pi\, G(z,-\lambda)}{|\ln(\lambda\bar\lambda)|+\ln(|E|)-4\pi/\alpha}    \right\}+ \nonumber\\
&+\frac{4i\, G(z,-\lambda)}{|\ln(\lambda\bar\lambda)|+\ln(|E|)-4\pi/\alpha}\sign(\lambda\bar\lambda-1)
\frac{\frac{i\sign(\lambda\bar\lambda-1)}{\lambda} a(\lambda)}{\frac{i}{\pi}a(\lambda) } =\label{eq:5.22} \\
&=\frac{i}{\lambda}\exp{\left[\frac{\sqrt{|E|}}{2}\left(\lambda\bar z + \frac{z}{\lambda}  \right)\right]} -\frac{i}{\lambda}\frac{4\pi\, G(z,-\lambda)}{|\ln(\lambda\bar\lambda)|+\ln(|E|)-4\pi/\alpha} +\nonumber\ \\
&+\frac{i}{\lambda}  \frac{4\pi\, G(z,-\lambda)}{|\ln(\lambda\bar\lambda)|+\ln(|E|)-4\pi/\alpha}\, 
\frac{ a(\lambda)}{a(\lambda) }=\frac{i}{\lambda}\exp{\left[\frac{\sqrt{|E|}}{2}\left(\lambda\bar z + \frac{z}{\lambda}  \right)\right]}.\nonumber\
\end{align}

This completes the proof of Theorem~\ref{thm:2}.\qed

\end{document}